\documentclass[11pt,preprint]{aastex}
\usepackage{graphicx}

\let\oldfootsep=\footnotesep
\newcommand\ltsima{$\; \buildrel <\over\sim \;$}
\newcommand\simlt{\lower.5ex\hbox{\ltsima}}
\newcommand\gtsima{$\; \buildrel >\over\sim \;$}
\newcommand\simgt{\lower.5ex\hbox{\gtsima}}

\newcommand\msun {M_\odot}

\newcommand\rsun { {R_\odot}}

\newcommand\mearth {{M_\oplus}}

\newcommand\pac{Paczy{\'n}ski }

\shorttitle{}
\shortauthors{Bennett, Anderson et al}

\begin{document}


\title{Identification of the OGLE-2003-BLG-235/MOA-2003-BLG-53\\
         Planetary Host Star\altaffilmark{1}}


\author{David~P.~Bennett\altaffilmark{2},
               Jay Anderson\altaffilmark{3},
               Ian A.~Bond\altaffilmark{4},
               Andrzej Udalski\altaffilmark{5},
               and Andrew Gould\altaffilmark{6}
       } 
\altaffiltext{1}{Based on observations made with the NASA/ESA Hubble 
Space Telescope obtained at the Space Telescope 
Science Institute under program GO-10426, which is operated by the 
Association of Universities for 
Research in Astronomy, Inc., under NASA contract NAS 5-26555.}

\altaffiltext{2}{Department of Physics,
    University of Notre Dame, IN 46556, USA\\
    Email: {\tt bennett@nd.edu}}

\altaffiltext{3}{Astronomy Department,
    Rice University, Houston, TX, USA\\
    Email: {\tt jay@eeyore.rice.edu} }
    
\altaffiltext{4}
{Institute of Information and Mathematical Sciences, Massey University,
Auckland, New Zealand,\\
Email: {\tt i.a.bond@massey.ac.nz}}

\altaffiltext{5}{Warsaw University Observatory,
    Al. Ujazdowskie 4
    00-478 Warszawa, Poland\\
    Email: {\tt udalski@astrouw.edu.pl}}

\altaffiltext{6}{
Department of Astronomy, The Ohio State University,
140 West 18th Avenue, Columbus, OH 43210\\
Email: {\tt gould@astronomy.ohio-state.edu}}


\clearpage

\begin{abstract}
We present the results of HST observations of the host star for the first
definitive extrasolar planet detected by microlensing. The light curve model for
this event predicts that the lens star should be separated from the source star
by $\sim 6\,$mas at the time of the HST images. If the lens star is a late G,
K or early M dwarf, then it will be visible in the HST images
as an additional source of light that is blended with the source image.
Unless the lens and
source have exactly the same colors, its presence will also be revealed by a
systematic shift between centroids of the source plus lens in different filter
bands. The HST data indicates both of these effects: the HST source that
matches the position of the source star is 0.21 magnitudes brighter in the
ACS/HRC-F814W filter than the microlensing model predicts, and there is an
offset of $\sim 0.7\,$mas
between the centroid of this source in the F814W and F435W
filter bands. We conclude the planetary host star has been detected in these
HST images, and this identification of the lens star enables a complete
solution of the lens system. The lens parameters are determined with a Bayesian
analysis, averaging over uncertainties in the measured parameters, interstellar
extinction, and allowing for the possibility of a binary companion to the
source star. This yields a stellar mass of
$M_\ast = 0.63{+0.07\atop -0.09}\msun$ and
a planet mass of $M_p = 2.6 {+0.8\atop -0.6} M_{\rm Jup}$
at an orbital separation of $4.3 {+2.5\atop -0.8}$AU.
Thus, the lens system resembles our own Solar System, with a
planet of $\sim 3$ Jupiter-masses in a Jupiter-like orbit
around a star of two-thirds of a Solar mass. 
These conclusions can be tested with future HST images, which
should reveal a broadening of the blended source-plus-lens
point spread function due to the relative lens-source proper motion.
\end{abstract}




\section{Introduction}
A new window on the study of extrasolar planets has been opened with
the discovery of four planets by the gravitational microlensing method
\citep{bond-moa53,ogle71,ogle390,ogle169}. This method
\citep{mao-pac,gouldloeb}
probes some regions of extrasolar planet parameter space that
are not accessible with other planet detection methods. For example,
microlensing probes a random sample of stars toward the Galactic
bulge and has no large selection effects based on stellar type. The
sensitivity of microlensing to low-mass planets 
\citep{em_planet,wamb}
at separations of a few AU
is also unique among current methods, as the recent discoveries of
OGLE-2005-BLG-390Lb and OGLE-2005-BLG-169Lb have demonstrated
\citep{ogle390,ogle169}. It is also relatively straight-forward to determine
the planet detection efficiency as a function of planet mass
\citep{planet-limit,gaudi-planet-lim}, and
the detection efficiency for such low-mass planets is more than an order
of magnitude lower than for Jupiter-mass planets. The microlensing
discovery of two low-mass planets, despite the much lower detection efficiency,
indicates that planets of $\sim 10\mearth$
are more common than Jupiter-mass planets at separations of a few
AU around the most common stars in our Galaxy, and the fraction
of stars with planets of $\sim 10\mearth$ is estimated to
be about 40\%. This would seem to 
confirm a key prediction of the core accretion model for planet
formation: that Jupiter mass planets are much more likely to
form in orbit around G and K-dwarf stars than around M-dwarfs 
\citep{laughlin,ida_lin,boss}.
In fact, \citet{laughlin} have argued that the Jupiter-mass planets found
in microlensing events \citep{bond-moa53,ogle71} are more likely to orbit 
white dwarfs than M-dwarfs. Clearly, the microlensing detections would
provide tighter constraints on planet formation theories if the properties of the host
stars were known.

In this paper, we present an analysis of images taken with the 
Hubble Space Telescope's Advanced Camera for Surveys (ACS)
of the source and lens stars for the first microlensing event
that yielded a definitive planet detection: 
OGLE-2003-BLG-235/MOA-2005-BLG-53. This event had a 
planetary lens system with a mass ratio of $q = 0.0039 {+0.0011\atop -0.0007}$,
so the planet is likely to have a mass similar to that of Jupiter or
larger. Like the three other definitive microlensing planet detections to
date, the light curve of this event has sharp features that resolve the
finite angular size of the source star. This implies that the features of
the lens system crossed the angular radius of the source star
in $t_\ast = 0.059\pm 0.007\,$days. This can be combined with the
source star angular radius, $\theta_\ast = 0.53\pm 0.04\,\mu$as
to yield the angular Einstein
radius: $\theta_E = t_E \theta_\ast /t_\ast = 0.55\pm 0.07\,$mas.
The angular source star radius 
has been determined from the source star magnitude and color
\citep{bond-moa53}
using the color-color relations of \citet{bessellbrett}
and the empirical relations between angular radius and surface V-K
brightness of \citet{vanbelle} and 
\citet{kervella_dwarf}. This corresponds to a linear source radius
of $1 \rsun$, for $D_S = 8.8\,$kpc, which is consistent with the
dereddened color $(V-I)_0 = 0.76$ of a G-dwarf source
\citep{bond-moa53}. Our $\theta_\ast$ value differs slightly from the value
given in \citet{bond-moa53} due to the improved zero-point
for ground-based photometry given below. $\theta_E$ is related to the
lens system mass by
\begin{equation}
M_L = {c^2\over 4G} \theta_E^2 {D_S D_L\over D_S - D_L} \ ,
\label{eq-m_thetaE}
\end{equation}
where $D_L$ and $D_S$ are the lens and source distances, and 
$M_L = M_\ast + M_p$ is the total lens mass. Since
$D_S$ is known (approximately) and
$M_\ast \gg M_p$, eq.~\ref{eq-m_thetaE} can be considered
to be a mass-distance relation for the lens star.

Eq.~\ref{eq-m_thetaE} can be combined with a multicolor main sequence star
mass-lumonosity relation \citep{kroupa_tout,gray,allen} to determine the brightness
and colors of the lens star as a function of its distance. This allows us to
predict the properties of the combined image of the lens and source in
the HST frames, as shown in Fig.~\ref{fig-predict}. The fraction of the combined
flux that is due to the lens, $f_{\rm lens}$, is shown in the top panel. 
During the microlensing event, the lens
and source were separated by $< 0.1\,$mas, but the light curve
indicates a relative proper motion of 
$\mu_{\rm rel} = \theta_\ast /t_\ast = 3.3\pm 0.4\,$mas/yr. So, the lens-source
separation was $5.9\pm 0.7\,$mas when the HST images were taken,
1.78 years after peak magnification, and color difference between the lens
and source stars implies that the centroid position for the unresolved
images of the lens and source will show a color dependence. This
color dependent centroid shift is shown in the bottom panel of Fig.~\ref{fig-predict}.
This figure indicates that
the lens star could contribute as much as $\sim 30$\% of the total lens+source flux
and yield a B-I centroid shift of $\sim 0.65\,$mas for a K-dwarf lens star of
0.6-0.8$\,\msun$.

\section{HST Data and Analysis}
\label{sec-hst}

\begin{figure}
\plotone{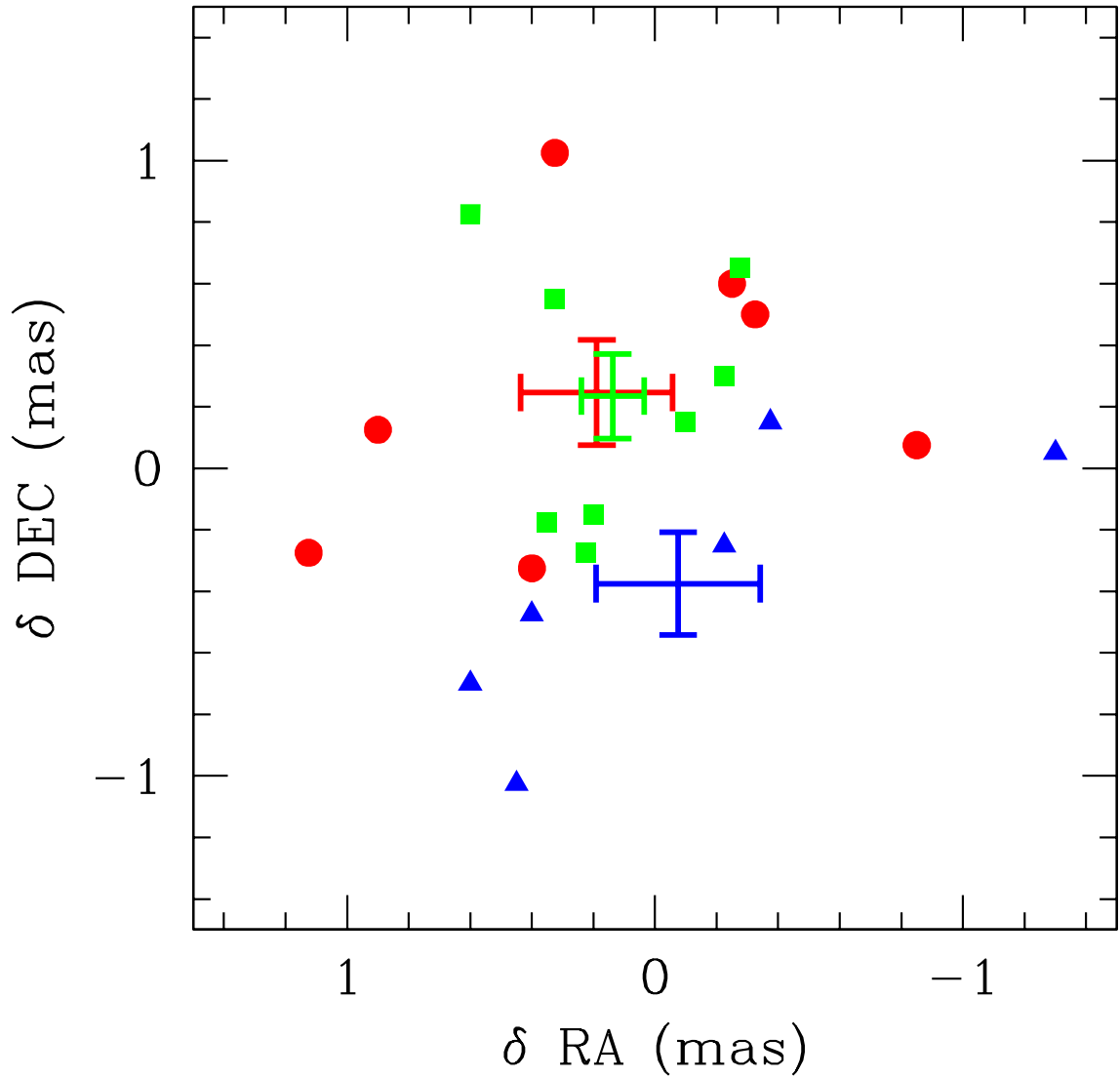}
\caption{
The centroids of the source$+$lens star blended images
in the individual HST/ACS/HRC images are shown as 
red circles (F814W), green squares (F555W), and
blue triangles (F435W). The crossed error bars are the
average centroid in each passband.
\label{fig-cent}}
\end{figure}

The HST data consist of 24 images with the ACS High Resolution Camera
(HRC) in 3 passbands: F814W ($8\times 60\,$sec exposures), F555W
($8\times 165\,$sec exposures), and F435W ($8\times 285\,$sec exposures).
The data were analyzed with the software program {\tt img2xym\_HRC}, 
documented in \citet{andking04} to measure positions 
and fluxes for all the possible sources in each of the 24 exposures 
with the appropriate library PSF for each passband. 
We then used the WCS header and 
centroid positions for stars in {\tt j96c01010\_drz.fits} to define 
a reference frame.  We used the 
crude positions in this frame for the initial 6-parameter linear
transformations from the distortion-corrected frame of each 
exposure into the reference frame.  Next, we combined the position 
for each star from the 24 exposures to get an average position 
for the star in the reference frame.  We then iterated the 
procedure, using the above average positions as the basis for 
the transformations, recomputing the average positions again.
We used these average positions as the basis for the final 
transformation from each exposure into the reference frame.  Each 
transformation was based on about 200 stars (excluding the 
target star) and used only stars 
with positions consistent to within 0.04 pixel, so that the 
transformations should be good to better than 0.003 pixel
($0.075\,$mas). This method was used to transform the position of 
the target star in each exposure into the reference frame,
as shown in Fig.~\ref{fig-cent}.

The HST photometry and astrometry were also compared directly to the
OGLE I-band data. The only
star in the HST frames within 0.3" of the source position is located
0.08" from the centroid of the difference image position of the source
star. This object in the HST frames must be the source star, but some
of the detected flux is likely to come from the lens star. This blended object
has an I-band (F814W) magnitude of $I_{\rm tot} = 19.42 \pm 0.02$,
whereas the microlensing fit indicates a source magnitude of
$I_S = 19.63\pm 0.06$ (in the VEGAMAG system), 
where the uncertainty includes both the
uncertainty in the microlensing fit and the relative normalization
of the OGLE and HST photometry. The $0.21\pm 0.06$ magnitude excess
seen in the HST I-band frames almost certainly comes from the lens
and/or binary companions to the lens or source star. If it 
comes from the lens, this would imply that 
$f_{\rm lens} = 0.18\pm 0.05$, which would imply that the lens must
be a main sequence star with 
$M_\ast = 0.64\pm 0.06 \msun$ according to the curve in Fig.~\ref{fig-predict}.

The bottom panel of Fig.~\ref{fig-predict} indicates that we should expect
color dependent centroid shifts of $\Delta r_{B-I} \sim 0.6\,$mas, 
$\Delta r_{V-I} \sim 0.4\,$mas, and $\Delta r_{B-V} \sim 0.2\,$mas.
The observed centroids are shown in Fig.~\ref{fig-cent}. (Two B-band
and one I-band images have been removed from this comparison due
to cosmic ray hits close to the target star, but including
these images would not change our conclusions.) While the scatter
between the individual measurements is larger than the difference
between the centroids from the different passbands, there does appear
to be a significant offset between the B-band and I-band centroids.
The measured average centroid shifts are
$\Delta r_{B-I} = 0.67\pm 0.26\,$mas, 
$\Delta r_{V-I} = 0.05\pm 0.26\,$mas, and
$\Delta r_{B-V} = 0.64\pm 0.23\,$mas (as shown in Fig.~\ref{fig-predict}). 
These error bars are based on the dispersion in the best fit centroids
in the individual images, and we have verified that the $\sim 20$ stars
of similar brightness without close neighbors near the center of the image 
follow have a centroid shift distribution centered at zero consistent
with their error bars.
The measured centroids of the target are consistent with
being colinear, but they are not independent. 
This $\Delta r_{B-I}$ value is consistent with expectations
for a main sequence lens star of mass 
$M_\ast = 0.64\pm 0.06 \msun$, while the $\Delta r_{V-I}$
and $\Delta r_{B-V}$ values are too small and too large by $\sim 1.5\,\sigma$,
respectively.

One complication to the interpretation of this event comes from the
parameters of a microlensing parallax fit to this event. We find an
apparent microlensing parallax signal that is significant at the $\sim 2\,\sigma$
level. This signal is too weak to precisely determine the direction of lens-source
relative proper motion, but motion in the NW or SW direction is preferred. This
contradicts the NNE direction indicated by the HST images. (The source is bluer
than the lens, so the lens moves up and slightly to the left in 
Fig.~\ref{fig-cent}). The parallax solution also prefers a smaller lens star
mass, so it would have difficulty accounting for the excess I-band flux
seen at the location of the source star. Thus, if we accept the parallax solution
at face value, then another star located within $\sim 10\,$mas of the source
may be required. This is unlikely to occur by chance because the
mean separation between stars in this field is $\sim 400\,$mas, so an
additional stellar companion to the source or lens star is the only reasonable
possibility. However, a companion to lens or source could give rise to
an additional acceleration (often called the xallarap effect) that can
easily mimic a weak microlensing parallax signal \citep{multi-par}. It is
also possible that a false microlensing parallax signal could be caused by small 
systematic photometry errors related to seasonal variations in airmass or
seeing. Because of these considerations, a 
potential microlensing parallax measurement is not included in our constraints.

\section{Bayesian Analysis}
\label{sec-bayes}

The HST data discussed in \S~\ref{sec-hst} and shown in Figs.~\ref{fig-predict}
and \ref{fig-cent} seem to favor a lens star mass of 
$M_\ast \approx 0.64\pm 0.06 \msun$, but excess flux and centroid
shift signals are in the 2.5-3.5$\,\sigma$ range. Also, there are a number
of uncertainties in the event parameters, such as the source distance, that
were ignored in Fig.~\ref{fig-predict}. The constraints from all measurements
can properly be included in a Bayesian analysis. We calculate the properties
of events based upon the mass function and
Galactic model used by \citet{gest-sim}, which employs
a double-exponential disk and a \citet{hangould-tau_stat} bar model for the
Galactic bulge. The source star is assumed to reside in the bulge, and the lens
star can reside in either the bulge or the disk. In order to account for the 
apparent microlensing parallax or xallarap signal, we assume that the source
star has a companion that is drawn from the same luminosity function that
is used for the lens stars. In this analysis, we marginalize over the source distance
and the $I_{\rm tot}$ and $E(V-I)$ measurements. For each set of lens, source
and source star companion magnitudes, we find the best fit source and lens
star centroids to match the measurements shown in Fig.~\ref{fig-cent}, subject to
the constraint on $\mu_{\rm rel}$.

We also allow variation in the assumed reddening and
extinction. Following \citet{bond-moa53}, we assume reddening of 
$E(V-I) = 0.82\pm 0.10$ for the source star,
and we convert this to extinction values in the different color bands using the
fitting formula of \citet{reddening} with $R_V = 2.1$ to account for the
anomalous extinction in the bulge 
\citep{udalski-extinct,ruffle-extinct,nishiyama-extinct}. Since the lens is in the
foreground of the source, the interstellar extinction for the lens star will be
less than for the source. The dust in the Galactic disk is thought to follow,
on average,
a double-exponential distribution with a scale height of about 100 pc
and the same scale length as the stellar distribution
\citep{drimmel}, but the distribution along any line-of-sight is quite
patchy. We model this uncertainty in the fraction of the total extinction
that lies in the foreground of the lens
with an exponential scale length of $1.5\pm 0.5\,$kpc, which
corresponds to a scale height of about $100\pm 50\,$pc in the direction of this
event. 
The variation in parameters such as $I_S$ and $E(V-I)$ affect the 
calculations that determine $\theta_\ast$, and $\theta_E$, and the
variation of these parameters is treated self-consistently.

\begin{figure}
\plotone{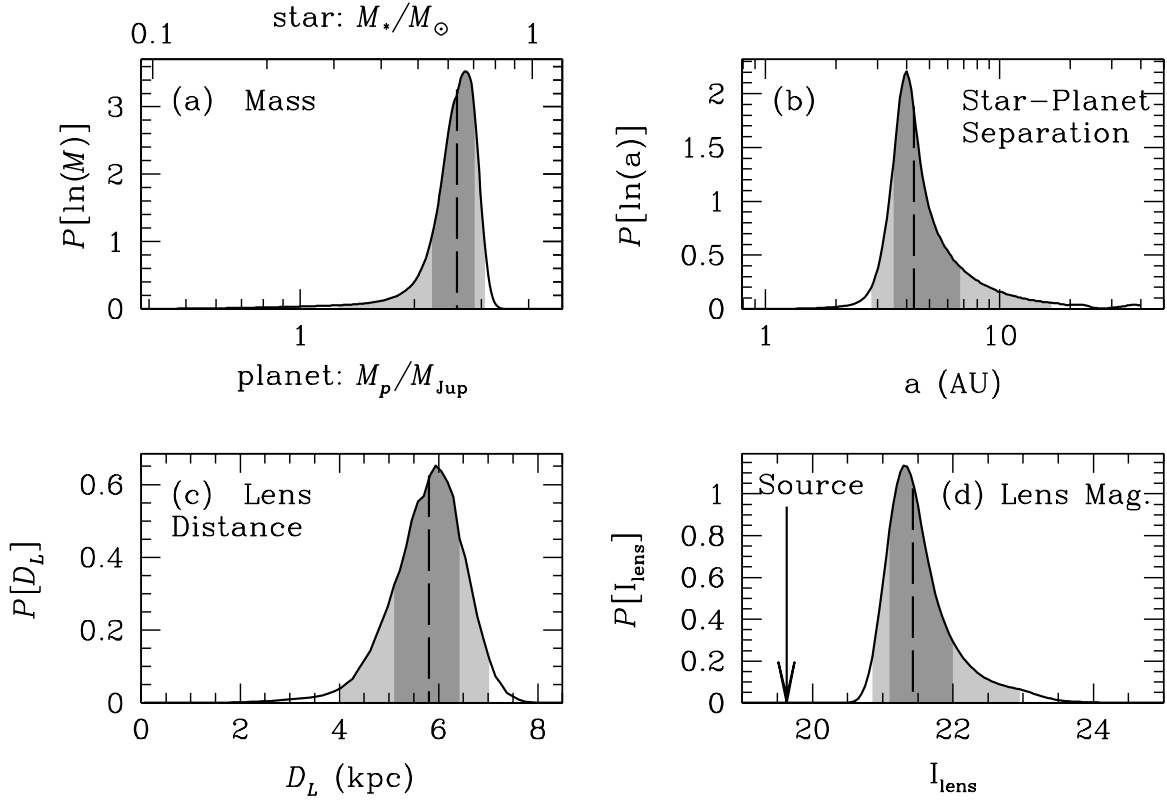}
\caption{
Bayesian probability densities for the properties of the planet
and its host star if it is a main sequence star. 
(a) The masses of the lens star and its planet ($M_\ast$
and $M_p$ respectively). 
(b) the separation, 
(c) their distance from the observer ($D_L$); 
and (d) the I-band brightness of the host star. 
The dashed vertical lines indicate the medians, and the shading
indicates the central 68.3\% and 95.4\% confidence intervals.  All
estimates follow from a Bayesian analysis assuming a standard model for the
disk and bulge population of the Milky Way, the stellar mass function of
Bennett \& Rhie (2002).
\label{fig-lens_prop}}
\end{figure}

The results of our Bayesian analysis are presented in Fig.~\ref{fig-lens_prop},
and the resulting median parameter values are 
$M_\ast = 0.63{+0.07\atop -0.09}\msun$,
$M_{\rm p} =  2.6 {+0.8\atop -0.6} M_{\rm Jup}$,
$a = 4.3 {+2.5\atop -0.8}\,$AU,
$D_L = 5.8 {+0.6\atop -0.7}\,$kpc, and
$I_L = 21.4 {+0.6\atop -0.3}$. These values are quite similar to the simple estimates
given in Sec.~\ref{sec-hst}, but the Bayesian analysis gives reasonable
estimates of the uncertainties. With the inclusion of the HST data, the
90\% c.l. range in lens star mass is 0.42-0.73$\,\msun$, which is
three times smaller than the allowed mass range without the HST data.
The mass range would be substantially smaller with a higher signal-to-noise
measurement of the centroid shift. (The excess brightness at the location
of the source could be primarily due to a binary companion to the source,
but a stronger detection of the centroid shift would exclude this possibility.)
We should also note that as a result of an error in the previous analysis
of the lens brightness constraint,
the lens star mass range presented here is not completely included by
the range presented in \citet{bond-moa53}.

\section{Discussion and Conclusions}

With the detection of the lens star for microlensing event
OGLE-2003-BLG-235/MOA-2005-BLG-53 with HST images
taken less than 2 years after peak magnification, we have
demonstrated that the ambiguities in the interpretation of
planetary microlensing light curves can readily be resolved with
high angular resolution follow-up observations. The resolution
of these ambiguities allows the planetary microlensing
events to place more significant constraints on planet formation
theories. For example, we can now confirm the core accretion
theory prediction of \citet{laughlin} that the 
OGLE-2003-BLG-235/MOA-2005-BLG-53 planetary host star is
not an M-dwarf. Instead, it is almost certainly a K-dwarf. If other
gas giant planets discovered by microlensing, such as
OGLE-2005-BLG-71Lb \citep{ogle71} also orbit stars that are
more massive than the typical microlens star, then this
would confirm the core accretion theory prediction that gas giant
planets form very rarely in orbit about low-mass stars 
\citep{ida_lin,laughlin}. However, the suggestion by \citet{laughlin} 
that the lens star might be a white dwarf is not supported by the 
HST data since a white dwarf lens would be much too faint to be
detected.

Although there is evidence that the lens star is detected in both
the observed flux from the target star and the color dependent
centroid shift, both of the signals are relatively weak. However, 
the centroid shift signal will grow with time, and will be 2.5 times
stronger if similar observations are made in late 2007. It should
also be possible to detect an additional effect: the broadening
of the target star PSF. In late 2007, the lens and source will
be separated by $\sim 14\,$mas, or about half an HST-ACS/HRC
pixel. According to detailed simulation by one of us (JA), an
exposure of  $\sim 2000\,$sec (about one orbit) should
suffice to measure the PSF broadening in the F814W passband.

\begin{figure}
\plotone{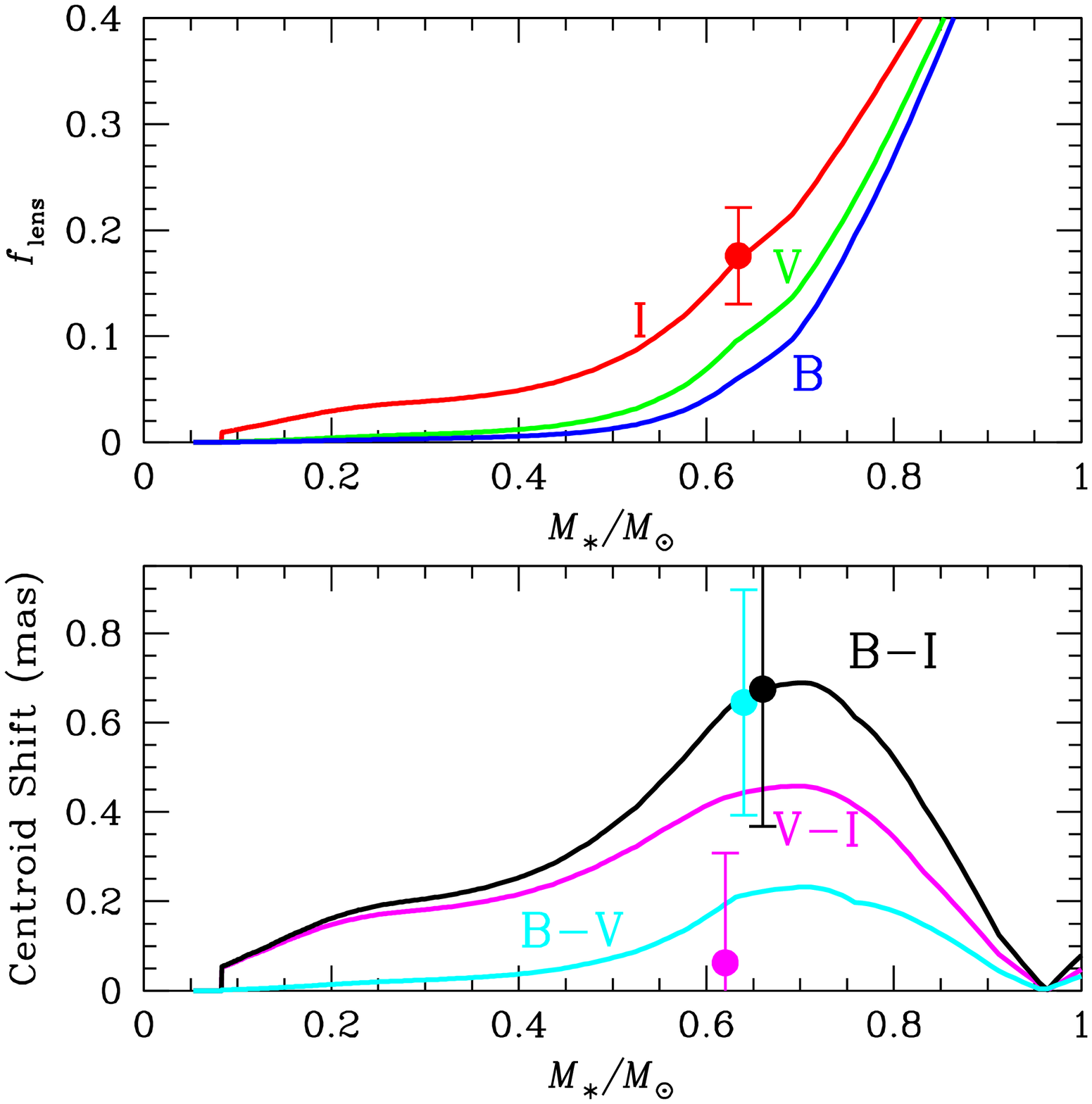}
\caption{
The top panel shows the fraction of the source$+$lens flux that is predicted to
come from the lens in the I (F814W), V (F555W), and B (F435W)
passbands as a function of lens mass. The bottom panel
shows the predicted color-dependent centroid shifts as a function
of mass for 1.78 years of relative proper motion at $\mu_{\rm rel} = 3.3\,$mas/yr.
The measured values of $f_{\rm lens}$ in the I-band and the color dependent
centroid shifts and error bars are indicated with their error bars. These
are plotted at stellar mass ($M_\ast$) values that were arbitrarily chosen to
be consistent with the analysis presented below.
\label{fig-predict}}
\end{figure}

It should be possible to identify the lens star for most other
planetary microlensing events. Because the relative proper motion,
$\mu_{\rm rel}$, is usually determined from light curve features for 
planetary microlensing events, we can routinely construct plots 
similar to Fig.~\ref{fig-predict} for other planetary 
events. The $\mu_{\rm rel}$ value for this event is smaller
than average, so this event is not particularly favorable for such
measurements. If the lens star had a lower mass, we might have
to wait $\sim 5$ years after the event or obtain more exposure time
to detect the centroid
shift. The situation for the host star of the $\sim 13\mearth$ planet
in event OGLE-2005-BLG-169, is more favorable because the
relative proper motion is large, $\mu_{\rm rel} = 8.4\pm 0.6\,$mas/yr,
and the source star is about a magnitude fainter. As a result,
the lens star can be detected in HST images 2 years after the
event, even if it is at the bottom of the main sequence (Bennett, 
Anderson \& Gaudi, in preparation). The only planetary microlensing event
to date that may not allow the detection of the lens star is
OGLE-2005-BLG-390 \citep{ogle390}, 
because the source star is a giant. However,
the lens star for this event may be detectable in the future with 
the Very Large Telescope Interferometer
(Beaulieu et al, in preparation). A space-based
microlensing survey \citep{gest-sim}, like the proposed
Microlensing Planet Finder \citep{mpf-spie} would routinely
detect the lens stars by the methods discussed in this
paper without the need for follow-up observations with a 
different telescope.

\acknowledgments
D.P.B.\ was supported by 
grants GO-10426 from the Space Telescope Science Institute (STScI),
AST 02-06189 from the NSF and NAG5-13042 from NASA. J.A.\ was
supported by the grant, GO-9443 from STScI, and A.G.\ was supported
by NSF grant AST 04-52758. 
A.U.\ was supported by the grant SP13/2003 of the Foundation for Polish
Science and MNiSW BST grant to the Warsaw University Observatory.

\clearpage


\clearpage


\clearpage


\end{document}